# HYDROGEN BOND VS π-STACKING INTERACTIONS IN THE *p*-AMINOPHENOL…*p*-CRESOL DIMER: AN EXPERIMENTAL AND THEORETICAL STUDY

M. C. Capello,[1] F. J. Hernández,[1] M. Broquier,[2,3] C. Dedonder-Lardeux,[4] C. Jouvet[4] and G. A. Pino[1]*

1) Instituto de Investigaciones en Físico Química de Córdoba (INFIQC) CONICET – UNC. Dpto. de Fisicoquímica – Facultad de Ciencias Químicas – Centro Láser de Ciencias Moleculares – Universidad Nacional de Córdoba, Ciudad Universitaria, X5000HUA Córdoba, Argentina.

2) Centre Laser de l'Université Paris Sud (CLUPS/LUMAT), Univ. Paris-Sud, CNRS, Institut d'Optique Graduate School, Univ. Paris-Saclay, F-91405 Orsay, France.

3) Institut des Sciences Moléculaires d'Orsay (ISMO), CNRS, Univ. Paris-Sud, Univ. Paris-Saclay, F-91405 Orsay, France.

4) Aix Marseille Université, CNRS, PIIM UMR 7345, 13397, Marseille, France.

* Corresponding author: gpino@fcq.unc.edu.ar

**ABSTRACT**

The gas phase structure and excited state lifetime of the *p*-AmPhenol...*p*-Cresol heterodimer have been investigated by REMPI and LIF spectroscopy with nanosecond laser pulses and pump-probe experiments with picosecond laser pulses, as a model system to study the competition between π-π and H-bonding interactions in aromatics dimers. The excitation is a broad and unstructured band. The excited state of the heterodimer is long lived (2.5 ± 0.5) ns with a very broad fluorescence spectrum red-shifted by 4000 cm$^{-1}$ with respect to the excitation spectrum. Calculations at the MP2/RI-CC2 and DFT-ωB97X-D levels indicate that Hydrogen-bonded (HB) and π-stacked isomers are almost isoenergetic in the ground state while in the excited state only the π-stacked isomer exists. This suggests that the HB isomer cannot be excited due to negligible Franck-Condon factors and therefore, the excitation spectrum is associated with the π-stacked isomer that reaches vibrationally excited states in the S$_1$ state upon vertical excitation. The excited state structure is an exciplex responsible for the fluorescence of the complex. Finally, a comparison was performed between the π-stacked structure observed for the *p*-Aminophenol...*p*-Cresol heterodimer and the HB structure reported for the (*p*-Cresol)$_2$ homodimer indicating that the differences are due to different optical properties (oscillator strengths and Franck-Condon factors) of the isomers of both dimers and not to the interactions involved in the ground state.

**INTRODUCTION**

Non-covalent interactions are very important in different areas of chemistry and molecular biology.[1,2] Particularly, π-π interactions and hydrogen-bonds (conventional and unconventional H-bond) between aromatics rings are associated with supramolecular structure and stability of biomolecules, such as proteins[3,4] and nucleic acids,[5,6] as well as important bio-recognition processes.[7-9] These non-covalent interactions, in aromatic dimers, can lead to different arrangements such as sandwich (S), parallel displaced (PD), tilted parallel-displaced (T-PD), T-shaped (T), tilted T-shaped (T-T), V-shaped (V) or hydrogen bond (HB) configurations, depending on the orientation of each ring.[10]

Recent theoretical results on benzene dimer, a benchmark system to understand π-π interactions, have shown that both, S and T structures are almost isoenergetic,[11-14] but only the T configuration has been observed experimentally.[15-17]

The most recent theoretical results on aromatics dimers are focusing on analyzing the substituent effects in the π-π interactions, studying complexes of benzene and substituted benzenes.[10,18-22] These studies revealed that all the dimers of substituted benzenes, in the S configuration, bind stronger than the benzene dimer, irrespective of the nature of the substituents, i.e., electron donating or withdrawing.[18,20] This result contradicts the Hunter-Sanders rules proposed to explain the substituent effects on the π-π interactions,[23] and therefore, the electrostatic term is not the sole factor governing the binding energy in this kind of interactions.

Many investigations focus on the competition between the different interactions that can occur in complexes of substituted aromatic molecules, especially to characterize the stabilization driving forces of π-stacked and HB dimers.

In some cases, the π-stacked structures have been characterized experimentally and assumed to be the most stable ones in the ground state of the dimer. Some of these π-stacked dimers involve N-heterocyclic aromatic rings[24-28] while there are a few cases in which π-stacking structures are formed by substituted benzene rings such as the aniline dimer[29], the 1,2-difluorobenzene dimer[30], the heterodimers aniline-benzene[31] and anisole-benzene[32] and more recently the homodimers of phenylacetylene[33] and anisole[34] have also been characterized as π-stacked structures.

In other cases, T, V and HB structures of (phenol)$_2$[35] and (*p*-cresol)$_2$[36] homodimers, and of 7-azaindole…fluoropyridines,[37] indole…pyridine,[38] indole…imidazole,[39] anisole…phenol,[40] and 7-azaindole…phenol[41] heterodimers have been experimentally observed. These structures are believed to be observed because of their remarkable stability in the ground state, as compared to other possible isomers.

In this work, we present an experimental and theoretical study of the *p*-aminophenol…*p*-cresol (*p*-AmPhOH…*p*-CreOH) heterodimer that offers the possibility of competition between many kinds of H-bond interactions between the different substituents (OH…OH, NH…OH, OH…NH) of both molecules as well as π-stacking interaction between the rings.

**EXPERIMENTAL**

The experimental set-up used in Córdoba for LIF and REMPI spectroscopy with nanosecond lasers has been described previously.[42] Briefly, the carrier gas He at 1.5 bar passed through two reservoirs, the first one containing *p*-CreOH at room temperature, and the second one containing *p*-AmPhOH heated up to 353-385 K. The mixture was co-expanded into a vacuum chamber through a 300 μm diameter pulsed nozzle (Solenoid General Valve, Series 9).

Both reactants, *p*-CreOH and *p*-AmPhOH, were purchased from Sigma-Aldrich and used without further purification.

LIF and REMPI spectra were recorded using a frequency doubled Sirah dye laser (FWHM = 0.08 cm$^{-1}$) operating with Rhodamine 590, Rhodamine 610, Rhodamine 640 and DCM, pumped by the second harmonic (532 nm) of a Nd:YAG laser (Quantel, Brilliant B, pulse duration: 6ns). For the REMPI experiments, the molecular beam was collimated by a skimmer and was crossed perpendicularly by the laser beam in the center of the extraction zone of a home-made Wiley-McLaren time-of-flight (TOF) mass spectrometer (MS) (46 cm flight length). The ions were extracted perpendicularly to the molecular beam and laser directions, and detected by a microchannel plate (Jordan MCP). For the LIF experiments, excitation (LE) and dispersion (DF), the jet was intercepted at right angle, by the laser beam, at 1.5 – 2.0 cm from the nozzle. The fluorescence was collected by a telescope collinear to the jet and detected by a photomultiplier tube (PMT) (Hammamatsu R636) without any filter, or dispersed by a monochromator (FWHM = 1 nm). The signals from PMT and MCP were averaged and digitized by a Tektronic (TDS-3034B) oscilloscope and integrated with a PC. The rise time of the complete detection system was 1 ns.

The experimental conditions used in Orsay were the same as in Córdoba. In this case, pump-probe experiments with picosecond laser pulses were performed. The molecular beam was crossed perpendicularly by the laser beams, 10 cm downstream from the nozzle, in the center of the extraction zone of a TOF-MS and the ions were accelerated toward a MCP detector located at the end of a 1.5 m field-free flight tube perpendicular to the jet and laser beams axis.

For the pump-probe experiments, the third harmonic (355 nm) output of a mode-locked picoseconds Nd:YAG laser (EKSPLA-SL300) was split in two parts to pump two OPA and SHG systems (EKSPLA-PG411) for obtaining tunable UV light. One of the systems was used as excitation laser tuned at several frequencies ($\nu_1$) while the other system was tuned to 325 nm and used as ionization laser ($\nu_2$), keeping its energy at 100 µJ/pulse approximately, while the energy of the $\nu_1$ laser was attenuated to preclude one-color two-photon ionization. The temporal shapes of both pulses were determined in the fitting procedure as Gaussian functions of (15 ± 2) ps FWHM,[43] while the spectral line width was 5 cm$^{-1}$. The laser pulses were optically delayed between -300 and 800 ps by a motorized stage.

**THEORETICAL CALCULATIONS**

*Ab-initio* calculations were performed with the TURBOMOLE program package,[44] making use of the resolution-of-the-identity (RI) approximation for the evaluation of the electron-repulsion integrals.[45] The equilibrium geometry of the clusters in their ground electronic state ($S_0$) was determined at MP2 level. The equilibrium geometry of the lowest excited singlet state ($S_1$) and the excitation energies were determined at the RI-CC2 level.[46] These calculations were performed with the correlation-consistent polarized valence double-zeta basis set (cc-pVDZ).[47] The Franck-Condon simulation was performed with the PGOPHER software[48] using the vibrational frequencies calculated for the ground and excited electronic states. Additionally, some faster DFT and TD-DFT calculations were performed with the GAUSSIAN 09 program package,[49] using the ωB97X-D functional[50] and the 6-311G++(d,p) basis set.

**RESULTS**

**Spectroscopy and excited state lifetime of *p*-AmPhOH…*p*-CreOH**

Figure 1 shows the one-color REMPI spectra, recorded with nanoseconds pulses, of *p*-AmPhOH, *p*-CreOH, (*p*-CreOH)$_2$ and *p*-AmPhOH…*p*-CreOH by integrating the intensity of the ions at m/z = 109, 108, 216 and 217 a.m.u., respectively. The spectra of the monomers show narrow and well defined transitions with the bands origin ($0_0^0$) for the S$_1$ ← S$_0$ transitions centered at 31395 cm$^{-1}$ for *p*-AmPhOH and 35336 cm$^{-1}$ for *p*-CreOH. The lowest energy part of the REMPI spectrum of (*p*-CreOH)$_2$ is also shown in the inset of Figure 1. The $0_0^0$ transition for this complex is red-shifted by - 324 cm$^{-1}$ from the $0_0^0$ transition of the free *p*-CreOH monomer as previously reported.[36] Under the present experimental conditions, well-defined progressions of low-energy vibronic modes are observed for this homodimer. The REMPI spectrum of the *p*-AmPhOH…*p*-CreOH complex, recorded with nanosecond pulses (green) and corrected by the dye laser power, is an unstructured broad band extending over 3000 cm$^{-1}$ (31000 and 34000 cm$^{-1}$), with the maximum at approximately 33000 cm$^{-1}$ and the apparent structure above 33500 cm$^{-1}$ is due to fluctuations of the laser power. The band origin ($0_0^0$) for the S$_1$ ← S$_0$ electronic transition cannot be determined in this case. The same broadband spectrum was obtained with picosecond pulses (orange trace).

The broad continuum spectrum may be due to a strong geometry change between the S$_0$ and S$_1$ states or to an ultra-short excited state lifetime of the complex or to hot complexes in the ground state that produce a congested spectrum.

The last possibility can be ruled out since we observed a structured spectrum for the (*p*-CreOH)$_2$, which indicates that the clusters formed in the jet are confined in low ro-vibrational levels of the electronic ground state.

To get more information on the excited state lifetime of the *p*-AmPhOH…*p*-CreOH complex, time resolved fluorescence (TR-LIF) experiments with nanosecond lasers as well as pump-probe ionization experiments with picosecond lasers were performed pumping at different excitation wavelength and probing at 325 nm for the latter case, which is enough to ionize the complex. The different values obtained by both methods are reported in Table I, together with the excited state lifetimes of the *p*-AmPhOH[51] and *p*-CreOH[52] monomers, reported previously by other authors.

The results show that the average excited state lifetime of the complex, determined by both techniques is (2.5 ± 0.5) ns, without any clear dependence on the excitation energy. Similar excited state lifetimes were reported for the *p*-AmPhOH (2.20 ns)[51] and *p*-CreOH (4.1 ns)[52] monomers, showing well resolved structured spectra. This result suggests that the lack of structure in the excitation spectrum of the heterodimer is not due to short excited state lifetime. In addition, structured spectra has been observed for short excited state lifetime species such as phenol…7-Azaindole dimer (30 ± 10 ps)[41] and *o*-aminophenol (35±5 ps).[53] Therefore, the broad unstructured excitation spectrum of the *p*-AmPhOH…*p*-CreOH complex is most likely due to a large geometry change between ground and excited states.

Finally, the dispersed fluorescence (DF) spectrum of the *p*-AmPhOH…*p*-CreOH complex, determined under the same experimental conditions as the REMPI spectrum, is shown in Figure 2. The DF spectrum (black), is a broad band extending over 8000 cm$^{-1}$ (32000 – 24000) cm$^{-1}$ with the threshold at about 24000 cm$^{-1}$ and the maximum of the band at about 29000 cm$^{-1}$.

**Geometry optimization and excitation spectrum simulation**

As mentioned above, the broad unstructured excitation spectrum of the *p*-AmPhOH…*p*-CreOH complex may be associated with a marked geometry change between ground and excited state structures of the dimer. Thus, the geometry of the dimer was optimized in both electronic states. The calculations were performed at the DFT-ωB97X-D and MP2 levels of theory for the $S_0$ state and TD-DFT-ωB97X-D and RI-CC2 levels for the $S_1$ state. The ωB97X-D functional is currently recommended to evaluate non-covalent complexes and a comparison of the results obtained from the MP2 and DFT methods (see below) supports this recommendation.[54]

A targeted exploration of the potential energy surface was carried out, and multiple trial structures with various bonding motifs, either HB or π-stacked, between both moieties were considered for the geometry optimization in the $S_0$ and $S_1$ state. The complete set of results is shown in Table S1 (supplementary information).

Overall, those structures in which the OH group from *p*-CreOH acts as H-donor and $NH_2$ group of *p*-AmPhOH as H-acceptor, leading to a HB and a π-stacked isomers, are the most stable at both theory level. Hereafter, we will work only on these two isomers since the others are not expected to be present in the molecular beam under the experimental conditions of this work (see Table S1 in S.I.).

The $S_0$ state relative energy and the vertical and adiabatic transition energies as well as the optimized structures in the $S_0$ and $S_1$ state for the HB and π-stacked isomers, calculated at both theory levels are shown in Table II. A good agreement between the results obtained from both theory levels, is observed from Table II, except for the relative stability of the HB and π-stacked isomers in the $S_0$ state. However, the energy difference is within the calculations error and therefore it is only an indication that both isomers are almost isoenergetic.

Quite remarkably, geometry optimization in the $S_1$ state, starting from HB or π-stacked ground state isomers, leads to the same π-stacked ($S_1$) structure.

For the HB isomer, a large geometry change between the HB($S_0$) and π-stacked($S_1$) is clearly observed (Table II) and negligible Franck-Condon factors are expected for its $S_1 \leftarrow S_0$ transition, so it might not be observed by optical excitation. For the π-stacked isomer, the geometry change between π-stacked($S_0$) and π-stacked($S_1$) is not as large as in the former case. Therefore, although both isomers could be present in the molecular beam, only the π-stacked can be observed.

A close inspection of the $S_0$ and $S_1$ state structures of the π-stacked isomer (Figure 3) shows that while both aromatic rings are displaced in the $S_0$ state, they are found quite aligned in the $S_1$ state. In addition, the inter-plane distance is shortened from 3.475 Å in the $S_0$ state to 2.932 Å in the $S_1$ state. Finally, the angle between the planes of the rings of both molecules decreases from ~7° to ~ 4.5 °. These geometry changes will induce a large activity of low frequency vibrations in the excitation spectrum and low Franck-Condon factors in the adiabatic excitation region.

The HOMO calculated at the geometry of the $S_0$ state is a π orbital whose electronic density is distributed almost equally on each ring in a way that allows the electrostatic interaction between them, while the LUMO calculated at the $S_1$ geometry is a bonding orbital with most of the electronic density between the rings (Figure 4).

Assuming that the π-stacked isomers is the only one that is optically active in the spectral region explored in this work, the excitation spectrum for the dimer was simulated using the calculated frequencies in the ground and excited states and the Franck-Condon factors computed by the Pgopher program[46] from the optimized geometries of the π-stacked($S_0$) and π-stacked($S_1$) at the RI-CC2 level. Figure 5 shows the simulated (red line) and experimental one-color REMPI ns (black line) spectra of the complex, together with the three most active vibrational modes, which are related to the more important geometry changes between the $S_0$ and $S_1$ states.

A good agreement between the experimental and the simulated spectrum of the π-stacked isomer is presented in Figure 5. The first band observed in the simulation does not correspond to the $0_0^0$ transition but to a higher vibrational level. The $0_0^0$ transition is not observed experimentally and according to the simulation it is expected to be at 3.76 eV (30328 cm$^{-1}$), red-shifted by 1067 cm$^{-1}$ from the $0_0^0$ transition of bare *p*-AmPhOH.

## DISCUSSION

The unstructured spectrum observed for the *p*-AmPhOH…*p*-CreOH complex is not due to spectral congestion associated with a bad cooling because the excitation spectrum of the related (*p*-CreOH)$_2$ complex, recorded under the same experimental conditions, is structured. Band broadening associated to a short excited state lifetime was also dismissed as responsible for the lack of structure since for this complex the measured lifetime is (2.5 ± 0.5) ns.

The experimental and theoretical evidence points to a π-stacked isomer that undergoes geometry changes upon electronic excitation. These geometry changes lead lo low Franck-Condon factors in the adiabatic excitation energy and then vibrationally excited states are reached in the $S_1$ state producing the unstructured spectrum with negligible intensity in the vicinity of the $0_0^0$ transition.

It is clear that the most important changes in geometry between $S_0$ and $S_1$ states are associated with the distance between the aromatic rings and their relative displacement, allowing a better overlap between the π clouds of both molecules in the $S_1$ state than in the $S_0$ state.

Moreover, as shown in Figure 4, the better overlap in the $S_1$ state induces that the electrostatic/dispersive interaction observed between the π clouds in the ground state becomes a bonding orbital with most of the electronic density between the rings in the $S_1$ state, suggesting a stronger interaction between the two aromatic molecules in the excited state, leading to an exciplex-like excited state. These results are in agreement with previous results of other authors on related systems in which broad excitation spectrum and red-shifted fluorescence spectrum were associated to the formation of exciplexes.[55-57]

**Comparison between *p*-AmPhOH…*p*-Cresol and (*p*-Cresol)$_2$**

While this work shows that the π-stacked isomer of the *p*-AmPhOH…*p*-Cresol complex is preferentially observed, the HB isomer was reported to be responsible for the structured excitation spectrum of the homodimer (*p*-Cresol)$_2$,[36] which is in line with the different character of the excitation spectra of these two complexes. These differences may be attributed to different relative stabilization energies of the isomers in each dimer. However, Table III shows that π-stacked and HB isomers are isoenergetic at the DFT-ωB97X-D and MP2 levels, in each complex, within the calculation error. Thus, it is expected that the HB as well as the π-stacked isomers of both dimers are present in the molecular beam and at similar concentrations. Therefore, this ground state property should not be the reason for which different isomers are observed in each complex.

Table III also shows a good agreement between the experimental transition energies and the calculated adiabatic $S_1 \leftarrow S_0$ transition energies for the assigned isomers. The computed oscillator strength at the TD-DFT (RI-CC2) levels for the HB 0.006 (0.05) and the π-stacked 0.006 (0.02) isomers of the *p*-AmPhOH…*p*-Cresol complex are very similar and they should be observed with the same probability by means of electronic spectroscopy. However, the extremely large change in geometry upon excitation of the HB isomers leads to negligible Franck-Condon factors for this transition and then the detection of this isomer becomes unlikely as previously observed in 7-Azaindol(H$_2$O)$_3$ complex.[43] For 7-Azaindol(H$_2$O)$_3$ complex, the most stable isomer was not observed due the large difference between the H-bonded network in the ground and excited state equilibrium geometries, that render very weak Franck-Condon factors, and the only isomer observed experimentally is found 0.3 eV above the most stable and unobserved one.

In the case of the (*p*-Cresol)$_2$ complex, the oscillator strength of the transition for the HB isomer 0.06 (0.04) is a factor of 200 (20) larger than the corresponding value for the π-stacked isomer 0.0003 (0.002) at the TD-DFT (RI-ADC2) level, which makes unlikely its detection by electronic spectroscopy. In the π-stacked homodimer, due to the exciton splitting, the first state is optically forbidden as observed from the calculations. The adiabatic transition energy for the first allowed state is found at 5.3 eV at the RI-ADC2 level with oscillator strength of 0.05 (RI-ADC2). However, this state is too high in energy to be responsible for the observed experimental transition (4.34 eV). This is assumed to be the reason for which only the HB isomer is observed for this complex.

For a long time many models were developed to explain the effect of the substitution on the preferential HB or π-stacking interaction in aromatic dimers and determine the forces involved.[10,18-22] We show here, that the remarkable difference observed upon substituting a CH$_3$ group by an NH$_2$ group in one of the *p*-Cresol molecules of the (*p*-Cresol)$_2$ dimer can be rationalized considering only the different detection probabilities for HB or π-stacking isomers, without invoking the forces involved in the ground state to stabilize one or the other. In fact, in the ground state both isomers are almost isoenergetic for the homo- and heterodimer.

**CONCLUSIONS**

The gas phase structure of the the *p*-AmPhOH…*p*-Cresol complex has been studied by REMPI, LIF and pump-probe experiments together with ab-initio and DFT calculations. From the results it is suggested that a π-stacked structure is responsible for the excitation spectrum. In the excited state this complex behaves as an exciplex-like. Almost isoenergetic with the π-stacked isomer, the HB isomer is not observed due to low Franck-Condon factors.

A comparison with the (*p*-Cresol)$_2$ dimer for which only the HB isomer was reported, indicates that the π-stacked isomer of this complex has an oscillator strength too low to be detected by electronic spectroscopy involving this excited state, although in this case, the HB and π-stacked isomers are also almost isoenergetic.

This is another example in which UV and/or IR-UV spectroscopy cannot be employed for searching the most stable structure in the ground state, since the reason for observing or not a given structure depends on the optical properties (oscillator strengths and Franck-Condon factors) of the transition and not only its ground state stability.

**ACKNOWLEDGMENTS**

This works was supported by ECOS-MinCyT cooperation program (A11E02), FONCyT, CONICET, SeCyT-UNC and the ANR Research Grant (ANR2010BLANC040501). We acknowledge the use of the computing facility cluster GMPCS of the LUMAT federation (FRLUMAT 2764). This research has been conducted within the international CNRS/CONICET laboratory LEMIR.

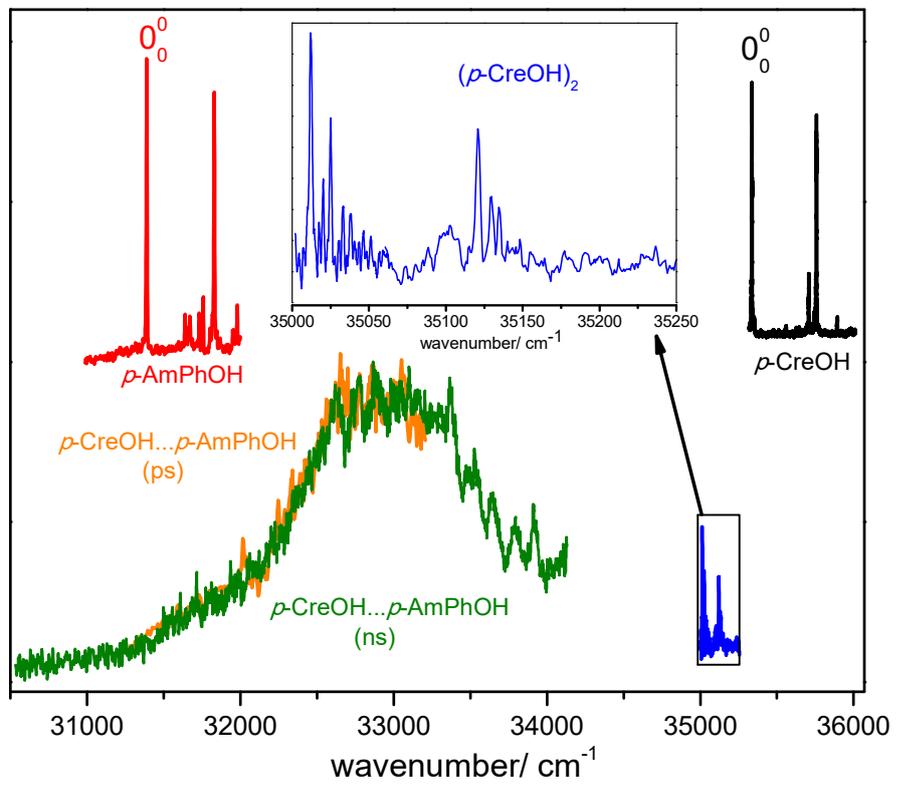

**Figure 1.** One-color REMPI spectra of *p*-AmPhOH (red trace), *p*-CreOH (black trace), (*p*-CreOH)$_2$ (blue trace) and *p*-AmPhOH...*p*-CreOH (green trace) recorded with nanosecond pulses. The orange trace corresponds to *p*-AmPhOH...*p*-CreOH REMPI recorded with picosecond laser. The inset shows an amplification of the (*p*-CreOH)$_2$ spectrum, where well defined transitions are observed.

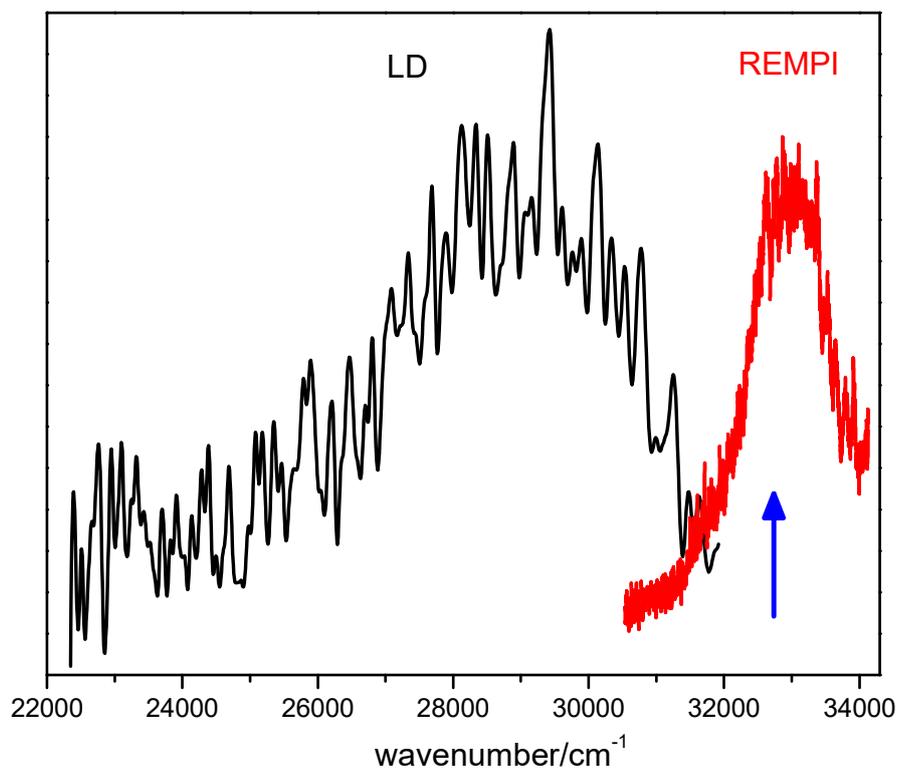

**Figure 2.** Dispersed fluorescence spectrum (DF, black trace) and excitation spectrum (red trace) of (*p*-AmPhOH...*p*-CreOH). The blue arrow indicates the excitation wavelength used to record the DF ( 32808.4 cm$^{-1}$).

**S₁**

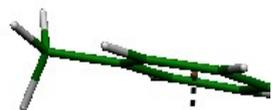 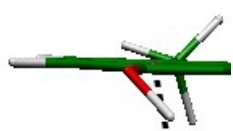 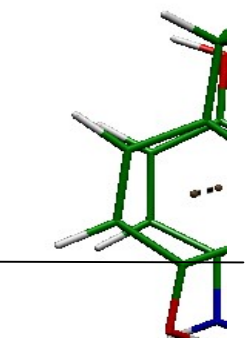

**S₀**

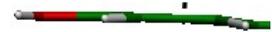 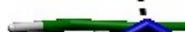

**Figure** 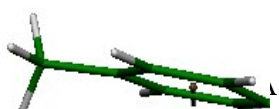 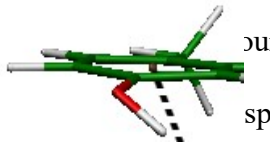 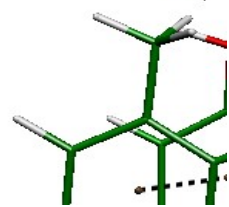

geometry between these states is clearly observed.

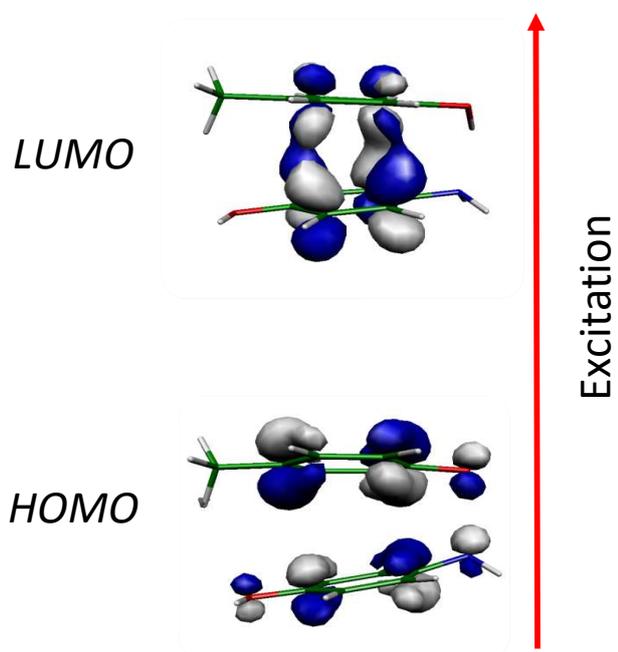

**Figure 4.** Orbitals involved in the S$_1$ ← S$_0$ electronic excitation for the *p*-AmPhOH...*p*-CreOH, calculated at the RI-CC2/cc-pVDZ level. In the excited state optimized geometry, an exciplex formation is observed.

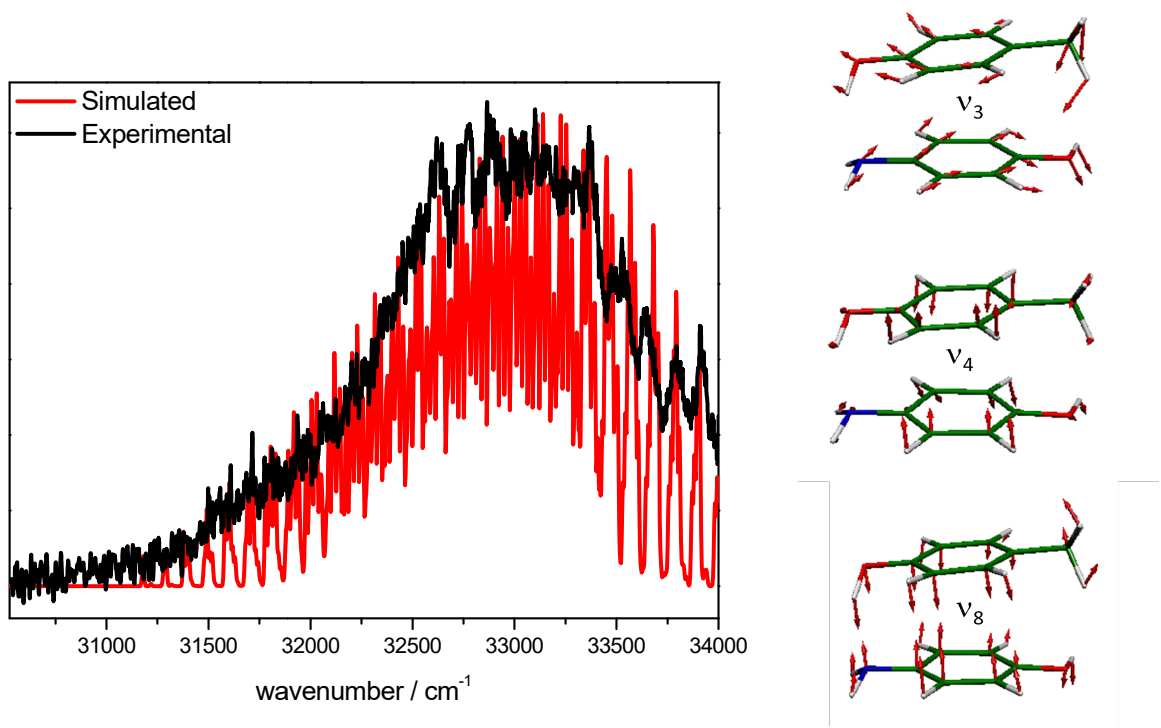

**Figure 5.** Left panel: Simulated spectra using the ground and excited frequencies calculated at RI-CC2/cc-pVDZ level for the *p*-AmPhOH...*p*-CreOH. In black, the experimental REMPI spectrum of the dimer recorded with nanosecond pulses. Right panel: Scheme of the most active vibrational modes in the spectrum.

**TABLE I.** Lifetimes of *p*-AmPhOH...*p*-CreOH and *p*-AmPhOH measured with picoseconds pump-probe experiments and resolved time-laser induced fluorescence (TR-LIF) at different excitation energy. The probe wavelength was 325 nm. The literature lifetime of the lowest energy transition of *p*-CreOH is showed.

[a] Ref. 51
[b] Ref. 52

| Complex or Molecule | Method | $\lambda_{excitation}$/ nm(cm$^{-1}$) | Lifetime / ns |
|---|---|---|---|
| *p*-AmPhOH...*p*-CreOH | ps pump-probe | 301.7 (33145.5) | (2.7 ± 0.3) |
| | ps pump-probe | 302.3 (33079) | (2.0 ± 0.6) |
| | ps pump-probe | 304.7 (32818) | (2.5 ± 0.3) |
| | TR-LIF | 303.2 (32981.5) | (2.9 ± 0.1) |
| | TR-LIF | 305.2 (32765.4) | (2.5 ± 0.1) |
| *p*-AmPhOH | - | 318.5 (31395) | (2.20 ± 0.03)[a] |
| *p*-CreOH | - | 282.99 (35337) | 4.1[b] |

**TABLE II.** Comparison between the optimized structures of *p*-AmPhOH...*p*-CreOH and energy difference (in eV) between the ground and excited states, calculated at the MP2/RI-CC2/cc-pVDZ and DFT/TD-DFT-ωB97X-D/6-311G++(d,p) levels of theories.

[a] ΔZPE was estimated from the MP2/RI-CC2 calculations

| | MP2/RI-CC2 | | DFT/TD-DFT-ωB97X-D | |
|---|---|---|---|---|
| | π-stack | HB | π-stack | HB |
| | Ground State Optimized Geometry | | | |
| | 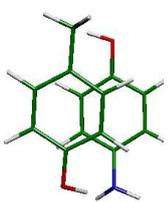 | 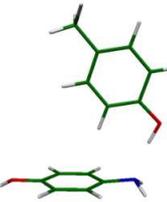 | 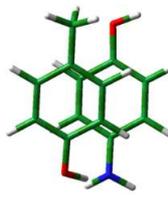 | 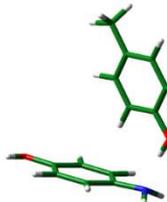 |
| Relative energy $S_0$-$S_1$ Vertical transition (without ΔZPE) | 4.50 | 4.60 | 4.64 | 4.70 |
| ΔZPE | 0.19 | - | 0.19[a] | 0.19[a] |
| $S_0$-$S_1$ Vertical transition (with ΔZPE) | 4.30 | - | 4.45 | 4.51 |
| | Excited State Optimized Geometry | | | |
| | 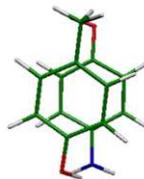 | 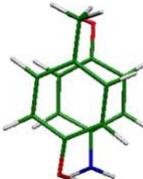 | 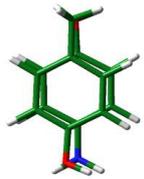 | 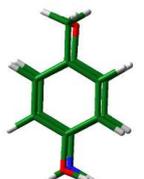 |
| $S_1$ | 3.95 | 3.88 | 4.06 | 4.05 |
| $S_0$-$S_1$ Adiabatic transition (with ΔZPE) | 3.76 | 3.69 | 3.87[a] | 3.86[a] |

**TABLE III.** Ground state relative energies, vertical and adiabatic $S_1 \leftarrow S_0$ transition energies (eV) and the corresponding oscillator strengths for the HB and π-stacked isomers of the *p*-AmPhOH…*p*-Cresol and (*p*-Cresol)$_2$ complexes, calculated at the DFT/TD-DFT-ωB97X-D/6-311G++(d,p) and MP2/RI-CC2/cc-pVDZ (values in parenthesis) levels of theories.

| Isomer | $S_0$ | $S_1$ vertical | $S_1$ adiabatic + $\Delta$ZPE[a] | Experimental | Osc. Strength |
|---|---|---|---|---|---|
| *p*-AmPhOH…*p*-Cresol | | | | | |
| HB | 0 (0.07) | 4.70 (4.60) | 3.86 (3.69) | - | 0.006 (0.05) |
| π-stacked | 0.02 (0.0) | 4.64 (4.50) | 3.87 (3.76) | ~ 3.76 | 0.006 (0.02) |
| (*p*-Cresol)$_2$ | | | | | |
| HB | 0.03 (0.0) | 5.01 (4.87) | 4.44 (4.46) | 4.34 | 0.06 (0.04) |
| π-stacked | 0.0 (0.02) | 4.97 (4.79) | 3.97 (3.92) | - | 0.0003 (0.002) |

[a] $\Delta$ZPE was estimated (0.19 eV) from the MP2/RI-CC2 calculations.